\documentclass[aps,pre,twocolumn]{revtex4}
\usepackage{graphicx}
\RequirePackage{times}
\usepackage[dvips,colorlinks,linkcolor=blue,citecolor=blue,urlcolor=blue]{hyperref}

\begin {document}
\title{ Discrete breathers in polyethylene chain}
\author{A.V. Savin}
\email{asavin@center.chph.ras.ru}
\author{L.I. Manevitch}
\email{lmanev@center.chph.ras.ru}
\affiliation{N. N. Semenov Institute of Chemical Physics,
Russian Academy of Sciences,
ul. Kosygina 4, 117977 Moscow, Russia}

\date{\today}

\begin{abstract}
The existence of discrete breathers (DBs), or intrinsic localized
modes (localized periodic oscillations of transzigzag) is shown.
In the localization region periodic contraction-extension of
valence C---C bonds occurs which is accompanied by
decrease-increase of valence angles. It is shown that the
breathers present in thermalized chain and their contribution
dependent on temperature has been revealed.
\end{abstract}
\maketitle

\section{Introduction}

Localized excitations in nonlinear systems (solitons, polarons,
breathers) became the subject of growing interest during last
decades. Discrete breathers, or intrinsic localized modes
correspond to shortwave length periodic vibrations which can not
be revealed in continuum model of the chain. Intensive study of
the breathers has an origin in  pioneer work of Sievers and Takeno
\cite{p1}. Their existence is proved by corresponding theorem
\cite{p2,p3} and numerous numerical investigations \cite{p4}. Now
their role in mechanisms of the energy relaxation in molecular
systems becomes rather clarified \cite{p5,p6,p7}.

However all these results relate to simple one-dimensional
systems. Meantime it should be waited that similar elementary
excitations can exist in more realistic discrete nonlinear models
\cite{p8}. The necessary conditions for their existence are
discreteness of the system leading to boundedness of phonon frequency
spectrum and amplitude dependence of eigenfrequencies caused by nonlinearity.
These conditions are satisfied for polymeric molecules. For example,
anharmonic potential of C---H valence bond in carbon-hydrogen
chains leads to existence of high-frequency localized vibrations
of such bonds \cite{p8}. Naturally, this type of breathers has to
be present in polyethylene (PE) chain but corresponding
frequencies are very high ($\sim 3100$ cm$^{-1}$). We will discuss
the breathers caused by the backbone vibrations of PE chains which
correspond to much more less frequencies because they arise due to
coordinated changes of C---C valence bonds and CCC-valence angles.

\section{Model of macromolecular chain in PE crystal}

Nonlinear dynamics of planar PE macromolecule has been considered
in \cite{p9,p10} where a detail consideration of the transzigzag
model is presented.
This is a reason for only short its description which is given below

\begin{figure}[thb]
\includegraphics[width=1.0\linewidth]{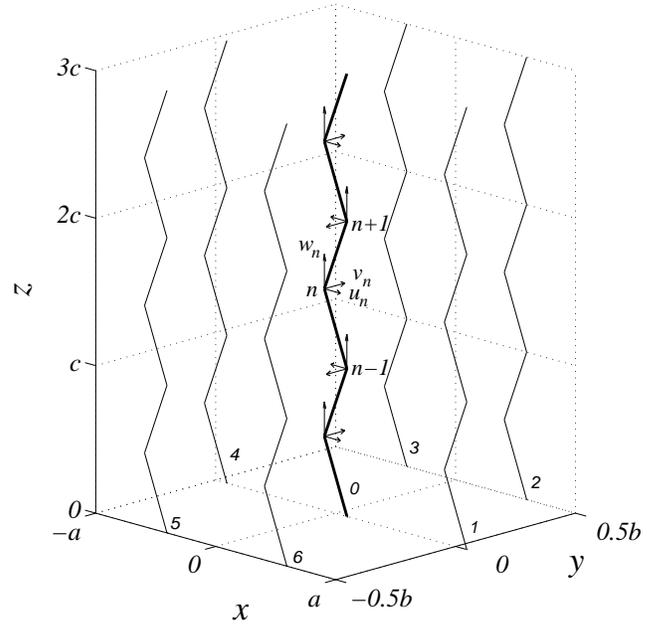}
\caption{\label{fig1}\protect
        Schematic representation of the crystalline PE. The
        central trans-zigzag backbone (curve 0) and the
        six neighbor chains (curves 1,2,...,6) are shown.
        The local coordinates for the central chain are presented.
        }
 \end{figure}

PE crystal consists of parallely packed zigzag chains [PE macromolecules
(CH$_2$)$_\infty$]. The structure of the crystal is schematically plotted
at Fig. \ref{fig1}. Every PE macromolecule  in the crystal has a
transzigzag conformation. It means that backbone of the chain has
a planar zigzag structure with equilibrium length $\rho_0=1.53$ \AA~~
of valence bond H$_2$C--CH$_2$ and value $\theta_0=113^\circ$ of valence
angle CH$_2$--CH$_2$--CH$_2$.
All macromolecules are situated in parallel planes. The crystalline structure
is characterized by angle $\theta_0$ and three periods: $a=4.51$ \AA,~ $b=7.031$ \AA,~
$c=2\rho_0\sin (\theta_0/2)=2.552$ \AA~.

Let us consider the dynamics of PE macromolecule and will take
into account its interaction with six neighbor chains which being
considered as immobile  (approximation of immobile neighbor chains).
Because the motion of hydrogen atoms with respect to backbone chain
is not significant when studying low energy nonlinear dynamics, we
will consider every CH$_2$-group as a united particle.

Let us introduce the local coordinate system presented at Fig.
\ref{fig1}. Then the Hamiltonian of the chain can be written as
follows
    \begin{eqnarray}
    H=\sum_n\{\frac12 M(\dot{u}_n^2+\dot{v}_n^2+\dot{w}_n^2)
   +V(\rho_n)+U(\theta_n) \nonumber \\
   +W(\delta_n)+Z(u_n,v_n,w_n)\},
    \label{f1}
    \end{eqnarray}
where the first term describes the kinetic energy of $n$th unit,
the second the deformation energy of the  $n$th valence bond,
the third  the deformation energy of the  $n$th valence angle,
the fourth  the deformation energy of $n$th torsional angle, and
the last term the energy of interaction of the $n$th unit with
the six neighboring  chains (substrate potential),  $M=14m_p$
is the mass of the united atom ($m_p$ is the proton mass).

The length of $n$th valence bond is
$$
\rho_n=(a_{n,1}^2+a_{n,2}^2+a_{n,3}^2)^{1/2},
$$
where $a_{n,1}=u_n+u_{n+1}-l_x$, $a_{n,2}=v_n+v_{n+1}$,
$a_{n,3}=w_{n+1}-w_n+l_z$ [$l_x=\rho_0\cos(\theta_0/2)$ and
$l_z=\rho_0\sin(\theta_0/2)$ -- transversal and longitudinal step
of the zigzag chain]. The cosine of the $n$-th valence angle is
$$
\cos(\theta_n)=\left(a_{n-1,1}a_{n,1}+a_{n-1,2}a_{n,2}
-a_{n-1,3}a_{n,3}\right)/\rho_{n-1}\rho_n,
$$
and the cosine of the $n$th torsional angle is
$$
\cos(\delta_n)=\left(-b_{n,1}b_{n+1,1}-b_{n,2}b_{n+1,2}+
b_{n,3}b_{n+1,3}\right)/\beta_n\beta_{n+1},
$$
where
\begin{eqnarray}
b_{n,1}&=&a_{n-1,2}a_{n,3}+a_{n,2}a_{n-1,3}, \nonumber \\
b_{n,2}&=&a_{n-1,1}a_{n,3}+a_{n,1}a_{n-1,3}, \nonumber \\
b_{n,3}&=&a_{n-1,2}a_{n,1}-a_{n,2}a_{n-1,1}, \nonumber \\
\beta_n&=&\left(b_{n,1}^2+b_{n,2}^2+b_{n,3}^2\right)^{1/2}. \nonumber
\end{eqnarray}

Potentials of valence bond, valence angle, and torsion angle are accepted
in the form
\begin{eqnarray}
V(\rho_n)&=&D_0\left\{1-\exp[-\alpha(\rho-\rho_0)]\right\}^2,
\nonumber \\
U(\theta_n)&=&\frac12\gamma(\cos\theta-\cos\theta_0)^2,
\nonumber \\
W(\delta_n)&=&C_1+C_2\cos\delta_n+C_3\cos 3\delta_n,
\nonumber
\end{eqnarray}
where the parameters $D_0=334.72$ kJ/mol, $\alpha=1.91$ \AA$^{-1}$,
$\gamma=130.122$ kJ/mol, $C_1=8.37$ kJ/mol, $C_2=1.675$ kJ/mol,
$C_3=6.695$ kJ/mol.  The substrate potential
\begin{eqnarray}
&Z(u,v,w)=\varepsilon_w\sin^2(\pi w/l_z)&\nonumber \\
&+\frac12K_u\left[1+\varepsilon_u \sin^2(\pi w/l_z)\right]
\left\{ u-\frac12l_x\left[1-\cos(\pi w/l_z)\right]\right\}^2&
\nonumber \\
&+\frac12K_v\left[1+\varepsilon_v\sin^2(\pi w/l_z)\right]v^2,&
\nonumber
\end{eqnarray}
where the parameters $\varepsilon_u=0.0674265$ kJ/mol,
$\varepsilon_v=0.0418353$ kJ/mol, $\varepsilon_w=0.1490124$
kJ/mol, $K_u=2.169513$ kJ/\AA~mol$^2$, $K_v=13.683865$
kJ/\AA~mol$^2$. Detail substantiation of chosen potentials
is presented in \cite{p10}.

\section{Linear oscillations of transzigzag}

Small-amplitude vibrations of isolated transzigzag were considered
in \cite{p11,p12,p9} and vibrations with account of interaction with
immobile surrounding chains -- in \cite{p10}.

\begin{figure}[tb]
\includegraphics[width=1.0\linewidth]{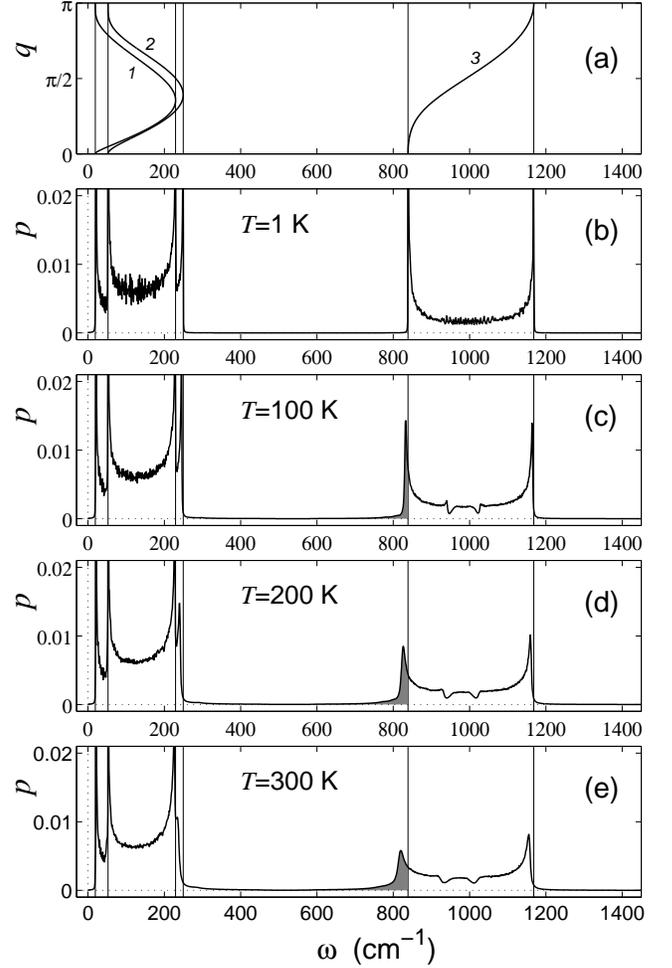}
\caption{\label{fig2}\protect
       Dispersion curves $\omega=\omega_a(q)$, $\omega=\omega_t(q)$,
       $\omega=\omega_o(q)$ (curves 1, 2, 3) for trans-zigzag interacting
       with immobile  surrounding chains (a).
       Density of energy distribution $p$ on frequencies $\omega$ for thermal
       vibrations with temperature $T=1$ K (b), $T=100$ K (c), $T=200$ K (d),
       and $T=300$ K (e). Gray color corresponds to frequencies region in
       which the discrete breathers occur.
       }
 \end{figure}

Small-amplitude vibrations can be divided on planar (in the zigzag plane)
and transversal ones. In turn, the plane motions are divided on
low-frequency acoustic and high-frequency optic vibrations.
Accordingly, one can separate three dispersion curves corresponding
to 1) plane acoustic phonons $\omega=\omega_a(q)$,
2) plane optic phonons $\omega=\omega_o(q)$, 3) transversal phonons
$\omega=\omega_t(q)$. These curves with account of interchain interaction
are presented at Fig. \ref{fig2} (a).

For isolated chain (substrate potential $Z(u,v,w)\equiv 0$) the acoustic phonons
have the frequency spectrum $0\le \omega_a\le 228.75$ cm$^{-1}$, torsional
(transversal) phonons -- spectrum $0\le \omega_t\le 243.85$ cm$^{-1}$, and
optic phonons -- spectrum $838.47$ cm$^{-1}\le \omega_o\le 1168.11$ cm$^{-1}$.
Interaction with surrounding chains leads only to a shift to right of low
boundary of acoustic and transversal phonons:
$19.05$ cm$^{-1}\le \omega_a\le 229.54$ cm$^{-1}$;
$52.43$ cm$^{-1}\le \omega_t\le 249.43$ cm$^{-1}$;
$838.69$ cm$^{-1}\le \omega_o\le 1168.27$ cm$^{-1}$.

\section{Localized nonlinear vibrations of PE chain}

The equations of motion corresponding to Hamiltonian (\ref{f1})
have the form
       \begin{eqnarray}
       M\ddot{u}_n=-\frac{\partial H}{\partial u_n},~~
       M\ddot{v}_n=-\frac{\partial H}{\partial v_n},~~
       M\ddot{w}_n=-\frac{\partial H}{\partial w_n},~~
       \label{f2}\\
       n=0,\pm 1,\pm 2,...~~. \nonumber
       \end{eqnarray}
Complexity of this system does not allow to find its analytic
solution. Thefore we used a numerical procedure.

The finite chain consisting of $N=200$ CH$_2$ groups was
considered. The viscous friction providing adsorption of phonons
was introdused on the boundaries of the chain.
The system of equations (\ref{f2}) with $n=1,2,...,N$
was integrated numerically with breather-like initial condition.
If a discrete breather can exist in the chain it can manifest
itself in irradiation of superfluous (non-breather part
of initial excitation) phonons.

\begin{figure}[tb]
\includegraphics[width=1.0\linewidth]{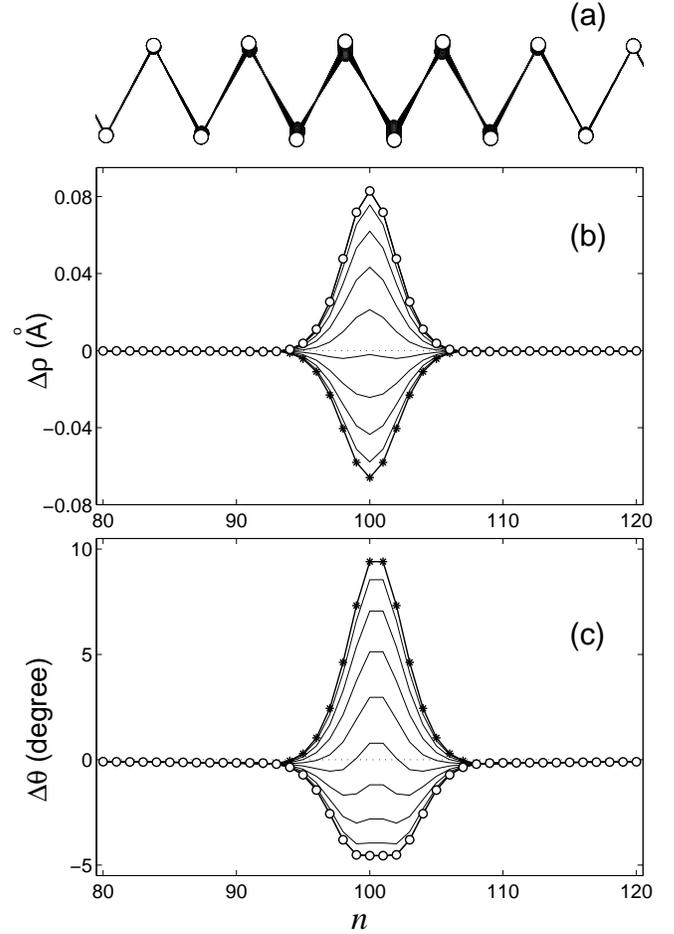}
\caption{\label{fig3}\protect
       Localized planar periodic vibrations of trans-zigzag .
       Vibrations are schematically shown, the thickness of
       line corresponds to amplitude (a). The magnitudes of
       valence bonds $\rho_n$ (b) and angles $\theta_n$ (c)
       are presented for ten different instants. The frequency
       of breather $\omega=820.5$ cm$^{-1}$, energy
       $E=26.4$ kJ/mol, width $L=4.28$
        }
 \end{figure}

Numerical modelling has shown that only one type of localized
periodic vibrations exists caused by tension-compression
of valence C---C bonds with coordinated change of valence angles
in the plane of transzigzag -- see Fig. \ref{fig3} (b) (c).
The vibration occurs in the plane
of trans-zigzag with nodes displacements perpendicular to
main backbone axis [Fig. \ref{fig3} (a)].

These vibrations are stable excitations, which are characterized by
frequency $\omega$, energy $E$ and dimensional width
$$
L=2\left[\sum_{n=1}^N (n-n_c)^2p_n\right]^{1/2},
$$
where the point $n_c=\sum_n np_n$ determines the position of the
vibrations center and sequence $p_n=\sum_n E_n/E$ -- the distribution
density of energy along the chain. Essential nonlinearity of these
vibrations is manifested in decrease of its frequency with amplitude
growth. So, the revealed excitation is actually discreet breather.

\begin{figure}[tb]
\includegraphics[width=1.0\linewidth]{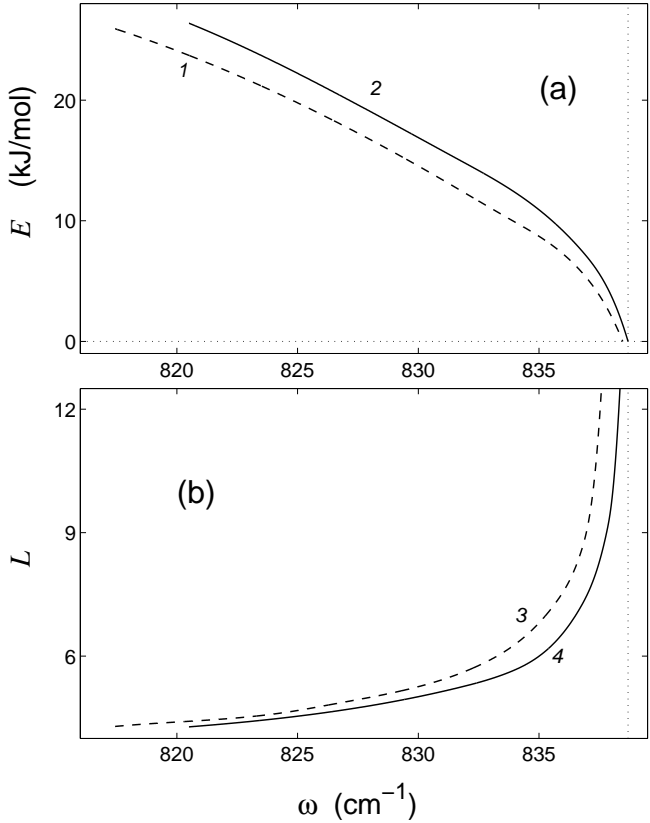}
\caption{\label{fig4}\protect
       Dependence of energy $E$ (a), width $L$ (b) of
       the breather upon frequency $\omega$ in isolated chain (curves 1, 3)
       and in the chain with substrate potential (curves 2, 4).
        }
 \end{figure}

Dependence of energy $E$ and width of the breather $L$ on its
frequency $\omega$ is presented at Fig. \ref{fig4}. The frequency
spectrum of the breather has situated near low boundary of optic
phonons. It is a reason why weak intermolecular interaction with
surrounding chains do not practically effect on such excitations.
If in isolated chain its frequency $\omega_b=817$
cm$^{-1}\le\omega<\omega_o(0)=838$ cm$^{-1}$, in the chain
surrounding by immobile neighbor chains
--- Ó$\omega_b=820$ cm$^{-1}\le\omega<\omega_o(0)=839$
cm$^{-1}$. When frequency decreases the energy of breather
increases monotonically and its width decreases monotonically.

The frequency of breather has to be separated from the frequencies
of small amplitude linear vibrations. One could expect also
the existence of low-frequencies breathers with frequencies near
upper boundary of transversal phonons $\omega>\max\omega_t(q)$ and
high-frequency breathers with frequencies near upper boundary
of optical phonons $\omega>\omega_o(\pi)$. But numerical analysis
has shown that such localized excitations are absent.

\section{Thermal vibrations of trans-zigzag as origin of discrete breathers}

Let us consider the thermal vibrations of trans-zigzag.
With this goal we analyze the finite chain consisting of N segments.
Their $N_0$ segments near boundary (from both sides) are situated into
heat bath with temperature $T$. The dynamics of the system is described
by the systems of Langevin equations
       \begin{eqnarray}
       M\ddot{u}_n=-\frac{\partial H}{\partial u_n}+\xi_n-\Gamma_nM\dot{u}_n,
       \nonumber\\
       M\ddot{v}_n=-\frac{\partial H}{\partial v_n}+\eta_n-\Gamma_nM\dot{u}_n,
       \label{f3}\\
       M\ddot{w}_n=-\frac{\partial H}{\partial w_n}+\zeta_n-\Gamma_nM\dot{u}_n,
       \nonumber \\
       n=1, 2,...,N~, \nonumber
       \end{eqnarray}
where the Hamiltonian of the system $H$ is given by Eq.
(\ref{f1}), $\xi_n$, $\eta_n$, and $\zeta_n$ are random normally
distributed forces describing the interaction of $n$th molecule with
a thermal bath, the coefficient of friction $\Gamma_n=0$ for
$N_0<n\le N-N_0$ and $\Gamma_n=\Gamma$ for $n\le N_0$ and
$N-N_0<n\le N$. Coefficient of friction $\Gamma=1/t_r$, where $t_r$
-- the relaxation time of the velocity of the molecule.
The random forces $\xi_n$, $\eta_n$, and $\zeta_n$ have the
correlation functions
       \begin{eqnarray}
       \langle\xi_n(t_1)\xi_m(t_2)\rangle=
       \langle\eta_n(t_1)\eta_m(t_2)\rangle=
       \langle\zeta_n(t_1)\zeta_m(t_2)\rangle= \nonumber \\
       2M\Gamma k_BT\delta_{nm}\delta(t_1-t_2), \nonumber \\
       \langle\xi_n(t_1)\eta_m(t_2)\rangle=
       \langle\xi_n(t_1)\zeta_m(t_2)\rangle=
       \langle\eta_n(t_1)\zeta_m(t_2)\rangle=0, \nonumber \\
       1\le n,m\le N_0,~~N-N_0<n,m\le N, \nonumber
       \end{eqnarray}
where $k_B$ is Boltzmann's constant and  $T$ is the
temperature of heat bath.

The system (\ref{f3}) was integrated numerically by the standard
forth-order Runge-Kutta method with a constant step of integration
$\Delta t$. Numerically, the delta function was represented as
$\delta(t)=0$ for $|t|>\Delta t/2$ and $\delta(t)=1/\Delta t$ for
$|t|\le\Delta t/2$, i.e. the step of numerical integration
corresponded to the correlation time of the random force.
In order to use the Langevin equation, it is necessary  that
$\Delta t \ll t_r$. Therefore we chose $\Delta t=0.001$ ps and
the relaxation time $t_r=0.1$ ps.

Let us consider a frequency distribution of kinetic energy of
thermal vibrations. For this goal the system (\ref{f3}) was
integrated numerically for $N=500$, $N_0=N/2$. While choosing the
initial conditions as corresponding to ground state of the chain
the system was integrated during $t=10t_r$ to bring it
in the thermal equilibrium. After that we calculated the density
of molecules kinetic energy distribution on frequencies $p(\omega)$.
To increase an accuracy, the density of distribution was calculated
using 1000 independent realizations of the chain thermalization.
The profile of the density of distribution for different values of
temperature is presented at Fig. \ref{fig2} (it is accepted that
$\int p(\omega ) d\omega=3$).

For temperature $T=1$ K the density of distribution
coincides practically with corresponding density for linearized system,
so anharmonicity is here not essential.
All vibrations are linear and only phonons are thermalized.
With further increasing of temperature the amplitude of thermal
vibrations also increases so their anharmonism is manifested.
For $T=100$ K  we see a shift of density behind the low boundary
of the spectrum of optic phonons which  becomes more pronounced
with further increase of the temperature. High frequency vibrations
in this region can be identified as breathers because they arise
in the frequency region $[\omega_b,\omega_o(0)]$, corresponding to
this type of excitations. The part of energy corresponding to
breathers may be found as
$$
p_b=\int_{\omega_b}^{\omega_o(0)}p(\omega)d\omega.
$$
The contribution of the breathers in thermal energy increases with
growth of temperature (for $T=1$ K it is $p_b=0.002$,
for $T=100$ K --- $p_b=0.106$), has a maximal value
$p_b=0.115$ for $T=200$ K and then decreases (for
$T=300$ K the breathers contribution corresponds to $p=0.083$).

\begin{figure}[t]
\includegraphics[width=1.0\linewidth]{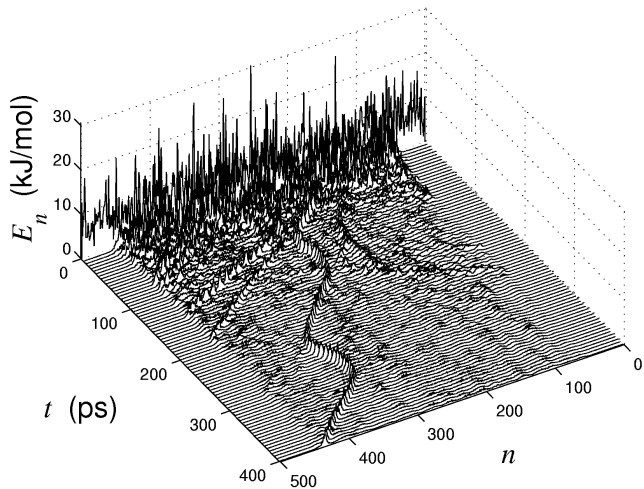}
\caption{\label{fig5}\protect
       Formation of discrete breathers from thermal vibrations of zigzag
       chain $(N=500,~~T=200$ K). The absorbing ends are considered ($N_0=50$).
       Temporal dependence of energy distribution $E_n$ in the chain is
       presented.
        }
 \end{figure}

Let us isolate the breathers from thermal vibrations.
With this goal we consider the chain consisting $N=500$ segments
with boundary cites $(N_0=50)$ embedded to the heat bath with temperature
$T$. After thermalization, we put the temperature of
thermal bath $T=0$ and consider the irradiation of heat energy from internal
region $(N_0<n<N-N_0)$. The relaxation process for $T=200$ K  is shown at Fig. \ref{fig5}.
We see a formation of several mobile localized excitations. Their detail analysis
leads to conclusion that they are discrete breathers with frequencies $\omega\sim\omega_b$.
So, one can observe the presence of breathers in thermal vibrations.

\begin{figure}[ht]
\includegraphics[width=1.0\linewidth]{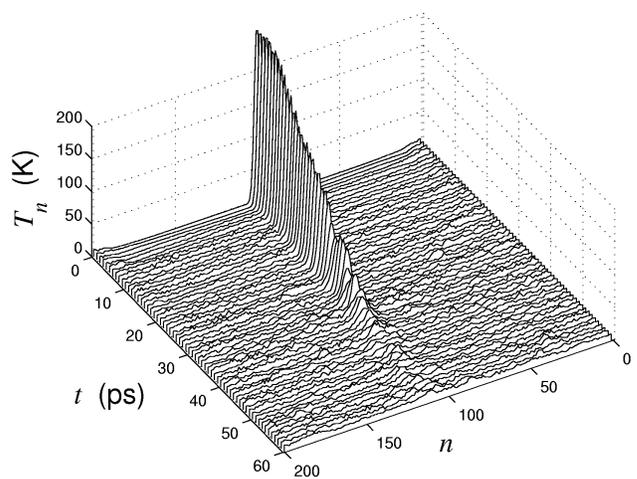}
\caption{\label{fig6}\protect
       Breaking  of the breather in thermalized chain
       ($N=200$, $N_0=10$, $T=10$ K, frequency of the breather
       $\omega=820.5$ cm$^{-1}$). Temporal dependence of current local
       magnitudes of temperature (kinetic energy of chain segments)
       $T_n$ is presented.
        }
 \end{figure}

Let us consider the interaction of discrete breather with thermal phonons.
The stationary discrete breather with frequency $\omega=820.5$
cm$^{-1}$ was situated in the center of finite chain
($N=200$), their edges ($N_0=10$) being situated in thermal bath
($T=10$ K). As we can see from Fig. \ref{fig6} the breaking of breather
is observed just as the center of the chain is thermalized
(the energy loss is 50\% for 20 ps).

Probability of thermally activated formation of discrete breathers in the chain
grows with increasing of the temperature. Therefore their concentration
has to increase when temperature grows. However in a thermalized chain
the breather has a finite time of life, decreasing with growth of
the temperature. It is reason for non-monotones dependence of the concentration
of breathers $p_b$ upon temperature $T$ --- it increases when
$T\le 200$ K and decreases for $T>200$ K with maximal magnitude for
$T=200$ K. Numerical study shows that the breathers may be better
separated from thermal vibration namely when $T=200$ K.

One can see from Fig. \ref{fig2} that revealed breathers is a unique type of
stable localized periodic excitations in thermalized chain
for given parameters of the crystal.
Besides the breathers only vibrations with frequencies possessing
to spectrum of linear oscillations
can be thermalized. It confirms the conclusions with respect
to existence of only one stable type of discrete breathers corresponding
to localized oscillations of valence C---C bonds.
The breathers present in thermalized
chain even for sufficiently small temperatures.

\section{Conclusion}

Stable localized nonlinear vibrations which are discrete breathers
can exist in polymer crystal. In PE macromolecule they are planar
vibration of transzigzag with periodic deformation of valence
bonds C---C and valence angles CCC. The breathers present in thermalized
chain and their contribution in heat capacity may be essential.

\end{document}